\documentclass[showpacs,twocolumn,preprintnumbers,amsmath,amssymb,prl,epsf,superscriptaddress]{revtex4-1}
\usepackage{graphicx}
\usepackage{nicefrac}
\usepackage[squaren]{SIunits}

\begin{document}
\title{Macroscopic Pure State of Light Free of Polarization Noise}
\author{Timur Sh.~Iskhakov}
\affiliation{Max Planck Institute for the Science of Light,
G\"unther-Scharowsky-Stra\ss{}e 1/Bau 24, 91058 Erlangen, Germany}
\author{Maria~V.~Chekhova}
\affiliation{Max Planck Institute for the Science of Light, G\"unther-Scharowsky-Stra\ss{}e 1/Bau 24, 91058
Erlangen, Germany} \affiliation{Department of Physics, M.V.Lomonosov Moscow State University, \\ Leninskie Gory,
119991 Moscow, Russia}
\author{Georgy~O.~Rytikov}
\affiliation{Department of Physics, M.V.Lomonosov Moscow State University, \\ Leninskie Gory, 119991 Moscow,
Russia}
\author{Gerd Leuchs}
\affiliation{Max Planck Institute for the Science of Light, G\"unther-Scharowsky-Stra\ss{}e 1/Bau 24, 91058
Erlangen, Germany} \affiliation{University of Erlangen-N\"urnberg, Staudtstrasse 7/B2, 91058 Erlangen, Germany}
\vspace{-10mm}

\begin{abstract}
The preparation of completely non-polarized light is seemingly easy: an everyday example is sunlight.
The task is much more difficult if light has to be in a pure quantum state, as required by most quantum-technology 
applications. The pure quantum states of light obtained so far are either polarized or, in rare cases, manifest
\textit{hidden polarization}: even if their intensities are invariant to polarization
transformations, higher-order moments are not. We experimentally demonstrate the preparation of the \textit{macroscopic singlet Bell state}, 
which is pure, completely non-polarized, and has no polarization noise.
Simultaneous fluctuation suppression in three Stokes
observables below the shot-noise limit is demonstrated, opening perspectives for noiseless polarization
measurements. The state is shown to be invariant to polarization transformations. This robust
highly entangled isotropic state promises to fuel important applications in photonic quantum technologies.
\end{abstract}
\vspace{5mm}
\maketitle \narrowtext

Polarization properties of light are at the focus of interest in quantum optics and quantum information. There
are many reasons, such as the relative simplicity of polarization interferometry and detection
schemes~\cite{Korolkova}, the similarity between the polarization properties of light and the properties of
atoms or atomic ensembles~\cite{Sorensen}, and several others. Description of polarization in quantum optics is
based on the Stokes operators $\hat{S}_0,\hat{S}_1,\hat{S}_2,\hat{S}_3$~\cite{Korolkova}.
Quantum noise in polarization observables is given by the
uncertainty relations
\begin{equation}
\Delta S_i\Delta S_j\ge|\langle S_k\rangle|, \,i\ne j\ne
k=1,2,3,\label{uncertainty}
\end{equation}
and it can be suppressed in $S_i$ at the expense of its increase in $S_j$ (polarization
squeezing~\cite{Chirkin,Bowen}).
In the particular case where $\langle S_k\rangle=0$, there are no fundamental restrictions for suppressing noise
in $S_{i}$ and $S_{j}$ simultaneously. Moreover, one can imagine a situation where the mean values of all three
Stokes operators, usually called the Stokes parameters, are equal to zero,
$\langle\hat{S}_1\rangle=\langle\hat{S}_2\rangle=\langle\hat{S}_3 \rangle=0$. This case can be realized for
entangled squeezed vacuum states~\cite{Braunstein,Karas,Karmas}.

Macroscopic, i.e., containing many photons, squeezed vacuum attracts much attention nowadays as it is among very
few `available' macroscopic quantum systems with essentially nonclassical properties~\cite{macroqubits}. In particular, it can be considered as a candidate for macroscopic Bell tests~\cite{Simon,Gisin}.
Besides, macroscopic squeezed vacuum has important applications such as gravitational wave detection~\cite{Lam},
quantum memory~\cite{Polzik}, super-resolution~\cite{Dowling}, quantum efficiency absolute
measurement~\cite{calibration}, etc. Entangled squeezed vacuum is a particular case of squeezed vacuum involving
at least four radiation modes, for instance, two polarization ones and two frequency ones. Its low-intensity
analogues are two-photon Bell states, and the preparation relies on the same experimental
scheme~\cite{Bell_Kul,Bell_Ryt}.

The preparation scheme involves two collinear nondegenerate optical parametric amplifiers (NOPAs) pumped by
coherent beams. The corresponding Hamiltonian takes one of the four possible forms,
\begin{eqnarray}
\hat{H}_{\psi\pm}=i\hbar G(a_{1}^{\dagger}b_{2}^{\dagger}\pm
b_{1}^{\dagger}a_{2}^{\dagger})+\hbox{h.c.},\nonumber\\
\hat{H}_{\varphi\pm}=i\hbar G(a_{1}^{\dagger}a_{2}^{\dagger}\pm
b_{1}^{\dagger}b_{2}^{\dagger})+\hbox{h.c.},
 \label{Ham}
\end{eqnarray}
where $a,a^{\dagger},b,b^{\dagger}$ are photon annihilation and creation operators in the horizontal and
vertical polarization modes, respectively, the subscripts $1,2$ denote two frequency modes, with the frequencies
$\omega_1,\omega_2$, and the parameter $G$ depends on the quadratic susceptibility and the pump power.

\begin{figure}[h]
\includegraphics[width=0.5\textwidth]{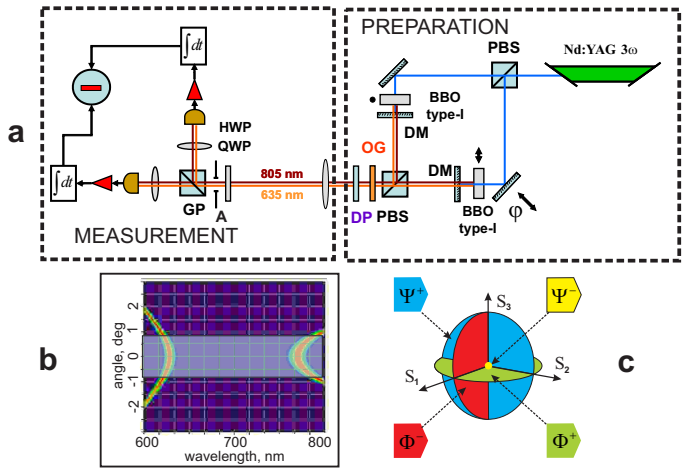} \caption{ \textbf{a}: the experimental setup. Orthogonally polarized squeezed vacuums are overlapped at a polarizing beamsplitter (PBS) with a relative phase
$\varphi$; the residual pump is eliminated by dichroic mirrors (DM) and a long-pass filter OG. Dichroic plate
(DP) is inserted for the production of $|\Psi^{(-)}_{mac}\rangle$. The measurement part includes a Glan prism
(GP), a half-wave plate (HWP), a quarter-wave plate (QWP), a lens, an aperture (A), and two detectors.
\textbf{b}: the wavelength-angular spectrum of PDC, the shaded area showing the range selected by the aperture.
\textbf{c}: macroscopic Bell states. Three coloured discs show the mean values and the uncertainties of the Stokes
observables for $|\Psi^{(+)}_{mac}\rangle$ (blue), $|\Phi^{(+)}_{mac}\rangle$ (green), and
$|\Phi^{(-)}_{mac}\rangle$ (red). The yellow point shows the state $|\Psi^{(-)}_{mac}\rangle$, for which the
means and uncertainties of all Stokes observables are zero.} \label{fig1}
\end{figure}

The states at the output can be written exactly as
\begin{eqnarray}
|\Psi^{(\pm)}_{mac}\rangle=e^{\Gamma(a_{1}^{\dagger} b_{2}^{\dagger}\pm
b_{1}^{\dagger}a_{2}^{\dagger})+\hbox{h.c.}}|\hbox{vac}\rangle,\nonumber\\
|\Phi^{(\pm)}_{mac}\rangle=e^{\Gamma(a_{1}^{\dagger} a_{2}^{\dagger}\pm
b_{1}^{\dagger}b_{2}^{\dagger})+\hbox{h.c.}}|\hbox{vac}\rangle, \label{state}
\end{eqnarray}
where $\Gamma=\int G\hbox{d}t$ is the parametric gain coefficient. Due to their similarity to the two-photon
Bell states, they can be called \textit{macroscopic Bell states}. The  Fock-state
expansion of (\ref{state}) immediately reveals photon-number correlation between modes $a_{1},b_{2}$ and between modes
$a_{2},b_{1}$~\cite{Simon,deMartini}. This leads to the suppression of fluctuations in certain Stokes
observables. However, for calculating observables such as Stokes parameters and their variances, it is more convenient to use the Heisenberg approach.
For instance, the Hamiltonian $H_{\psi -}$ leads to the equations of motion
\begin{equation}
\dot{a}_{1}=Gb_{2}^{\dagger},\,\dot{b}_{2}=Ga_{1}^{\dagger},\,
\dot{a}_{2}=-Gb_{1}^{\dagger},\,\dot{b}_{1}=-Ga_{2}^{\dagger},\nonumber
\end{equation}
the dot denoting time differentiation. The solutions are
\begin{eqnarray}
a_{1}=a_{10}\cosh\Gamma+b^{\dagger}_{20}\sinh\Gamma,b_{2}=b_{20}\cosh\Gamma+a^{\dagger}_{10}\sinh\Gamma,\nonumber\\
a_{2}=a_{20}\cosh\Gamma-b^{\dagger}_{10}\sinh\Gamma,b_{1}=b_{10}\cosh\Gamma-a^{\dagger}_{20}\sinh\Gamma,\nonumber
\end{eqnarray}
where `$0$' labels vacuum operators. The mean values and the variances of the Stokes operators are then calculated by averaging the operators $\hat{S}_i$ and $(\hat{S}_i-\langle\hat{S}_i\rangle)^2$, respectively, over the vacuum state. Note
that the Stokes operators for the output beam should be calculated as the sum of the Stokes operators for the two
wavelengths, $\hat{S}_i=\hat{S}_i^{(1)}+\hat{S}_i^{(2)}$~\cite{Simon,Karas}, where $\hat{S}_i^{(j)}$ is the $i-th$ Stokes operator for mode $j$; $i=0,1,2,3; j=1,2$. This is because the
detectors measuring intensities in various polarization modes (Fig.~\ref{fig1}a), even if fast enough to trace
intensity fluctuations, are still too slow to measure the beats caused by the $\omega_1-\omega_2$ difference.

For the Stokes parameters, we get
\begin{equation}
 \langle
S_0\rangle=4\sinh^2\Gamma,\langle S_1\rangle=\langle S_2\rangle=\langle S_3\rangle=0 \label{Stokes_par}
\end{equation}
for all states (\ref{state}), which clearly shows that they are unpolarized in the first order in the intensity.
It is different, however, for higher-order moments. For the variances, $\hbox{Var}S_i\equiv\langle \hat{S}_i^2\rangle-\langle \hat{S}_i\rangle^2\equiv\Delta S_i^2$, we get the results shown in Table 1. We see
that for each of the macroscopic Bell states, variance of at least one of the Stokes operators is zero, i.e.,
noise in the corresponding Stokes observable is completely suppressed~\cite{Karas,Karmas}.
\begin{table}[h]
\caption{Variances of the Stokes observables for the four states (\ref{state}); $n\equiv\sinh^2\Gamma$. Each
state has fluctuations in some Stokes observable suppressed; the $|\Psi^{(-)}_{mac}\rangle$ state has no
fluctuations in all Stokes observables $S_{1,2,3}$.}\label{table1}
\begin{center}
\begin{tabular}{|p{0.15\linewidth}|p{0.2\linewidth}p{0.2\linewidth}p{0.2\linewidth}p{0.2\linewidth}|}

\hline State  & $\hbox{Var}(S_0)$ & $\hbox{Var}(S_1)$ & $\hbox{Var}(S_2)$ & $\hbox{Var}(S_3)$\\
\hline
$|\Psi^{(-)}_{mac}\rangle$          & $8n(n+1)$    & $0$               & $0$               & $0$\\
$|\Psi^{(+)}_{mac}\rangle$          & $8n(n+1)$    & $0$               & $8n(n+1)$    & $8n(n+1)$\\
$|\Phi^{(-)}_{mac}\rangle$          & $8n(n+1)$    & $8n(n+1)$    & $0$               & $8n(n+1)$\\
$|\Phi^{(+)}_{mac}\rangle$          & $8n(n+1)$    & $8n(n+1)$               & $8n(n+1)$    & $0$\\
\hline
\end{tabular}
\end{center}
\end{table}

The \textit{macroscopic singlet state}, $|\Psi^{(-)}_{mac}\rangle$, is special. It follows from the structure of
the Hamiltonian $\hat{H}_{\psi-}$ that $|\Psi^{(-)}_{mac}\rangle$ is invariant to all polarization
transformations, similarly to the two-photon singlet Bell state. For this reason, it was called
\textit{polarization-scalar light}~\cite{Karmas}; unlike the other three states (\ref{state}), it does not reveal
polarization even in higher orders in the intensity (hidden polarization, \cite{Klyshko,Usachev,SS}). At the
same time, it is not a mixed state but a pure one. Even more surprising is the fact that the macroscopic singlet
state is completely noiseless from the polarization viewpoint. Indeed, according to Table~\ref{table1}, it has
the variances of the three Stokes observables $S_{1,2,3}$ exactly equal to zero. Moreover, one can show that higher-order moments
of these observables are zero as well. Note that this state, predicted theoretically as early as in the 1990-s~\cite{Karas,Karmas}, has never been observed before.

The noise properties of the macroscopic Bell states can be illustrated by a diagram shown in Fig.~\ref{fig1}c.
For each state, the coloured distribution is centered around the point ${\langle S_1\rangle;\langle S_2\rangle;\langle
S_3\rangle}$, while the size of each distribution in direction $S_i$ corresponds to the uncertainty $\Delta S_i$. For
the `triplet' states $|\Psi^{(+)}_{mac}\rangle, |\Phi^{(+)}_{mac}\rangle,|\Phi^{(-)}_{mac}\rangle$, the distributions
have the shapes of discs, as the uncertainty is zero in only one direction. For $|\Psi^{(-)}_{mac}\rangle$, the
distribution is pointlike, as all $\Delta S_i=0$. Formally, these distributions can be described by a
quasiprobability function~\cite{Karmasexp} introduced in Ref.~\cite{Wolf,polWig}, similarly to the way
distributions of optical quadratures can be described by the Wigner function.

The degree of noise suppression for a Stokes observable $S_i$ can be characterized by its normalized variance,
$\hbox{Var}(S_i)/\langle S_0\rangle$. Normalized this way, the variance turns into the noise reduction factor
(NRF) for the beams in two orthogonal polarization modes (horizontal and vertical for $i=1$, linear $\pm45\deg$
for $i=2$, and right- and left-circular for $i=3$). NRF, defined as the variance of the intensity difference for
two beams normalized to their mean intensity sum~\cite{Aytur}, is a parameter commonly used to quantify
twin-beam squeezing. The value NRF=1, often called the shot-noise level, is realized for the case of two
coherent beams and sets the boundary between the classical and quantum behaviour. Although theoretically NRF can
reach zero values (see Table~\ref{table1}), it is never the case in a real setup, because of the finite quantum
efficiencies, inevitable optical losses, and mismatch of signal and idler mode selection~\cite{two-color}. If
all losses are incorporated into some effective quantum efficiency $\eta$, equal for both detectors, the zeros
in Table~\ref{table1} turn into $4n(1-\eta)\equiv(1-\eta)\langle S_0\rangle$. Of course, losses also influence
the anti-squeezed variances of the Stokes observables, which become $4n(1+\eta)+8n^2\eta\equiv(1+\eta)\langle
S_0\rangle+\eta\langle S_0\rangle^2/2$. The corresponding minimum and maximum values of NRF are, respectively, $1-\eta$ and $1+\eta
+\eta\langle S_0\rangle/2$.

In our experiment (Fig.~\ref{fig1}a), we generate the macroscopic singlet state $|\Psi^{(-)}_{mac}\rangle$ via
frequency-nondegenerate parametric down-conversion (PDC) in two 2mm BBO crystals with optic axes oriented in orthogonal
(horizontal and vertical) planes placed into a Mach-Zehnder interferometer (MZI) whose input and output beamsplitters are polarizing ones. The crystals are pumped by Nd:YAG laser third harmonic (wavelength $\lambda_p=355$nm, repetition rate $1$ kHz, pulse duration $17$ ps, energy per pulse up to $0.2$ mJ). The pump at the input of the MZI is $45^{\circ}$
polarized and hence it contributes equally to PDC in both crystals.  After the crystals, the pump radiation is cut off by dichroic
mirrors reflecting $99.5\%$ of the pump and transmitting $95\%$ of the PDC radiation. Further suppression of the residual pump radiation is made by means of a color-glass OG filter. Each crystal is a traveling-wave NOPA producing a two-color bright squeezed vacuum~\cite{two-color} with signal and idler wavelengths $\lambda_1=635$ nm and $\lambda_2=805$ nm (Fig.~\ref{fig1}b). Both squeezed-vacuum beams, being
orthogonally polarized, leave through the same port of the MZI. The phase $\varphi$ between the
 squeezed-vacuum beams can be varied by moving one of the mirrors, placed on a piezoelectric feed. Depending on the phase, the states $|\Phi^{(+)}_{mac}\rangle$ or $|\Phi^{(-)}_{mac}\rangle$ are generated at the output of the MZ interferometer.

Preparation of the singlet state proceeds then along the same line as in Ref.~\cite{Bell_Ryt}. As the first step, we fix the phase to be $\pi$,
which is controlled by measuring the variance of $S_2$: the minimal value of the $S_2$ variance indicates the preparation of the
$|\Phi^{(-)}_{mac}\rangle$ state (see Fig.~\ref{fig1}c). In a $45^{\circ}$-rotated basis, the $|\Phi^{(-)}_{mac}\rangle$ state becomes
$|\Psi^{(+)}_{mac}\rangle$~\cite{Leuchs}. Finally, the $|\Psi^{(+)}_{mac}\rangle$ state is transformed into the $|\Psi^{(-)}_{mac}\rangle$ state. This is done with the help of a dichroic waveplate placed into the output beam.  The dichroic waveplate is a
$170\mu$ thick quartz crystal with the optic axis oriented at $45^{\circ}$. Its thickness is chosen so that the
o-e phase delays introduced at the wavelengths $\lambda_1$ and $\lambda_2$ differ by exactly $\pi$. The plate introduces a $\pi$ phase between the $a^{\dagger}_1b^{\dagger}_2$ and $b^{\dagger}_1a^{\dagger}_2$ terms in the expression for
$|\Psi^{(+)}_{mac}\rangle$ (see (\ref{state})), and therefore provides the necessary transformation from  $|\Psi^{(+)}_{mac}\rangle$ to  $|\Psi^{(-)}_{mac}\rangle$.

The registration part of the setup (Fig.~\ref{fig1}a, left), including a Glan prism, a half-wave plate (HWP) and a quarter-wave plate (QWP),
and two detectors, provides a standard Stokes measurement. With a HWP oriented at $22.5^{\circ}$, the
difference of detectors' output signals corresponds to $S_2$, while with a QWP oriented at $45^{\circ}$, the
same measurement provides $S_3$. Intermediate directions in the space of the Stokes variables are accessed by continuously rotating the HWP or QWP. Mode selection is performed by means of a lens with the focal length $300$ mm, placed at $300$ mm from the crystals, and a $10$-mm
aperture, placed in its focal plane. This way we select a $1.8^{\circ}$ angular spectrum width (Fig.~\ref{fig1}b), which
automatically restricts the PDC frequency spectrum as well. All optical elements are anti-reflection coated. The radiation is focused on the detectors by means of lenses with focal lengths $5$ cm. The detectors are pin photodiodes followed by charge-sensitive amplifiers, and their output pulses have a fixed shape (with the duration about $8\,\mu$s) and the amplitudes scaling as the number of photons registered during a single pump pulse. The electronic noise is mostly caused by the amplification circuit and is equivalent to $180$ input photons. For each pump pulse, the output pulses of the detectors are integrated in time by means of an AD card; then the data are processed to obtain the mean values and variances of the Stokes parameters. The shot-noise level is measured separately using attenuated laser radiation. In more detail, the procedure for measuring the SNR is described in Ref.~\cite{Iskhakov}. However, unlike in Ref.~\cite{Iskhakov}, in our present experiment the squeezed-vacuum pulses contain, on the average, more than $10^6$ photons, hence the shot noise exceeds the electronic noise by an order of magnitude.

\begin{figure}[h]
\begin{center}
\includegraphics[width=0.5\textwidth]{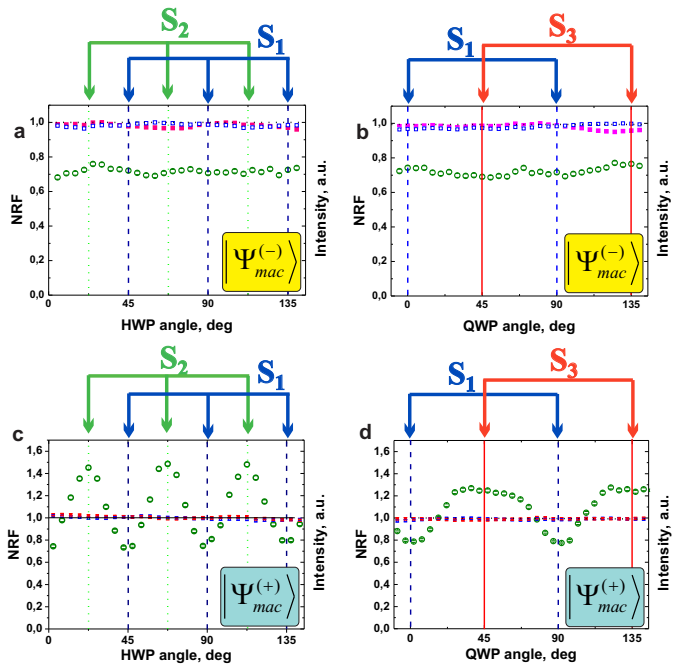}
\caption{Polarization properties of the macroscopic singlet state $|\Psi^{(-)}_{mac}\rangle$ (\textbf{a,b}) and
another macroscopic Bell state, $|\Psi^{(+)}_{mac}\rangle$ (\textbf{c,d}). Green large empty circles: NRF of the
Stokes observable versus the orientation of the HWP (\textbf{a,c}) and QWP (\textbf{b,d}). Blue dashed lines
mark the measurement of $\hbox{Var}(S_1)/\langle S_0\rangle$, green dotted lines, of $\hbox{Var}(S_2)/\langle
S_0\rangle$, and red solid lines, of $\hbox{Var}(S_3)/\langle S_0\rangle$. Small squares
denote normalized signals in both detectors.}\label{fig2}
\end{center}
\end{figure}

Using this setup, we have studied the polarization properties of the macroscopic singlet state.  We measured NRF for $S_1$, $S_2$, $S_3$, as well as the
intermediate Stokes observables corresponding to arbitrary orientations of HWP and QWP (Fig.~\ref{fig2}a,b, respectively).
Certain points in the obtained dependences correspond to the measurement of noise reduction in the Stokes
observables $S_1$ (Fig.~\ref{fig2}a,b, blue dashed lines), $S_2$ (Fig.~\ref{fig2}a, green dotted lines) and
$S_3$ (Fig.~\ref{fig2}b, red solid lines). For more clarity, arrows on top of each figure indicate the measurement of $S_1$, $S_2$, $S_3$. We see that all three observables $S_{1,2,3}$ have fluctuations
suppressed about $30\%$ below the shot-noise level. In particular, we have measured $NRF(S_1)=0.72\pm
0.01,\,NRF(S_2)=0.72\pm 0.01,\,NRF(S_3)=0.73\pm 0.02$. This moderate noise suppression, despite the overall
quantum efficiency/transmittivity of the setup being $\eta\approx 0.65$, is mainly due to the fact that both
$635$nm and $805$nm beams are restricted by the same angular aperture: for proper mode matching, the apertures
for different wavelengths should differ in size~\cite{two-color}.

In addition to noise reduction, the data clearly show the invariance of the singlet state to polarization
transformations with HWP and QWP. The normalized Stokes variance (large green open circles), as well as the intensities in the two output
channels (small red and blue squares), do not depend on the orientations of the plates. This justifies the term `polarization-scalar
light'. The state is a pure one but, at the same time, completely unpolarized. For comparison, we have also
tested the polarization properties of $|\Psi^{(+)}_{mac}\rangle$, which was generated the same way as
$|\Psi^{(-)}_{mac}\rangle$ but without the dichroic phase plate. The results are shown in Fig.~\ref{fig2}c,d.
One can see that although the output intensities are independent of the polarization rotation, the normalized
variances show well-pronounced modulation, which proves that the state has nonzero second-order degree of
polarization~\cite{Klyshko,Vik}. The other two macroscopic Bell states also reveal this \textit{hidden polarization} effect, but these results deserve a separate consideration and will be published elsewhere~\cite{to_be}. In the context of this work, it is important that the singlet state is  polarized neither in the first order in the intensity nor in the second one. 

Our results demonstrate the existence of pure intense unpolarized state of light with noise suppressed in all
polarization observables. One can say that this state \textit{manifests the strongest possible nonclassical
correlation between the intensities in orthogonal polarization modes regardless of the choice of these modes}.
This makes macroscopic singlet Bell state a candidate for macroscopic Bell tests and other challenging
fundamental experiments. Apart from its fundamental interest, this state will certainly find important
applications in quantum technologies based on the interaction between light and matter.


We acknowledge the financial support of the European Union under project COMPAS No.
212008 (FP7-ICT) and of the Russian Foundation for Basic Research, grants 10-02-00202 and 08-02-00741.
T.Sh.I. acknowledges funding from Alexander von Humboldt Foundation. We are grateful to I.~N.~ Agafonov for his
help in constructing the MZI.

\end{document}